\documentclass[useAMS]{mn2e}
\usepackage{lscape,graphicx}
\pdfoutput=1
\newcommand{\apg}{\:^{>}_{\sim}\:}
\newcommand{\apl}{\:^{<}_{\sim}\:}

\newcommand{\etal}{et al.}

\newcommand{\kms}{\mbox{km\ s${^{-1}}$}}
\newcommand{\lya}{\mbox{${\rm Ly}\alpha$}}

\begin{document}


\title[The Unchanging CGM Over 11 Billion Years]{The Unchanging Circumgalactic Medium Over the Past 11 Billion Years}
\author[Chen]{Hsiao-Wen Chen$^{1}$\thanks{E-mail:hchen@oddjob.uchicago.edu} \\
\\
$^{1}$Department of Astronomy \& Astrophysics, and Kavli Institute for Cosmological Physics, University of Chicago, Chicago IL 60637 }


\maketitle

\label{firstpage}

\begin{abstract}

  This paper examines how the circumgalactic medium (CGM) evolves as a
  function of time by comparing results from different absorption-line
  surveys that have been conducted in the vicinities of galaxies at
  different redshifts.  Despite very different star formation
  properties of the galaxies considered in these separate studies and
  different intergalactic radiation fields at redshifts between
  $z\approx 2.2$ and $z\sim 0$, I show that both the spatial extent
  and mean absorption equivalent width of the CGM around galaxies of
  comparable mass have changed little over this cosmic time interval.

\end{abstract}


\section{INTRODUCTION}

It is understood that galaxy growth is regulated by three competing
processes: mergers, gas accretion that supplies the fuel necessary to
support star formation, and feedback that returns gas and heavy
elements to intergalactic space and consequently disrupts continuing
gas accretion (Ceverino \& Klypin 2009).  Which process dominates and
over what period of time determines not only the properties of the
general galaxy population in different epochs, but also the chemical
enrichment history of the intergalactic medium (IGM).  Circumgalactic
space provides a critical laboratory for investigating the interplay
between star-forming regions and the IGM.

A powerful means to study and characterize the circumgalactic medium
(CGM) is to carry out absorption spectroscopy of background QSOs in
the vicinities of distant galaxies.  By searching for absorption
features in the spectrum of the background QSO that are coincident in
redshift with known galaxies in the foreground, empirical constraints
on the CGM such as absorber strength and gas kinematics can be
obtained at the projected separations between the galaxy and QSO pairs
(e.g.\ Lanzetta \etal\ 1995; Bowen \etal\ 1995; Steidel \etal\ 2002).
Unlike 21~cm observations which offer detailed two-dimensional maps of
extended H\,I gas around individual galaxies, each QSO sightline
provides only one measurement per galaxy.  However, observations of a
large sample of QSO-galaxy pairs with different projected separations
can give a two-dimensional map of the CGM averaged over the entire
galaxy sample.

Significant progress has been made over the past decade in mapping the
CGM out to $\sim 300$ kpc projected distances at redshift $z\apl 0.5$
using a large sample of QSO-galaxy pairs.  This includes the spatial
distributions of tenuous gas probed by the H\,I $\lya\,1215$
absorption transition (e.g.\ Chen \etal\ 1998, 2001a; Tripp \etal\
1998; Prochaska \etal\ 2011), chemically-enriched cool ($T\sim 10^4$
K) clouds probed by the Mg\,II $\lambda\lambda\,2796, 2803$ absorption
doublet (e.g.\ Chen \& Tinker 2008; Kacprzak \etal\ 2008; Barton \&
Cooke 2009; Chen \etal\ 2010a,b; Gauthier \etal\ 2010; Gauthier \&
Chen 2011), chemically-enriched ionized gas probed by the C\,IV
$\lambda\lambda\,1548, 1550$ absorption doublet, (e.g.\ Chen \etal\
2001b), and highly ionized warm gas probed by the O\,VI
$\lambda\lambda\,1031, 1037$ absorption doublet (e.g.\ Chen \&
Mulchaey 2009; Prochaska \etal\ 2011; Tumlinson \etal\ 2011).
However, studies of the CGM using QSO absorption spectroscopy become
significantly more difficult with increasing redshift, both because
the surface density of bright QSOs is low and because high-redshift
galaxies are faint (for example, an $L^*$ galaxy has ${\cal R}\approx
24.5$ mag at $z\sim 3$; e.g.\ Adelberger \& Steidel 2000) and
moderate-resolution spectroscopy of these faint galaxies requires a
large amount of telescope resources (e.g.\ Steidel \etal\ 2004; Erb
\etal\ 2003; Simcoe \etal\ 2006).

At the same time, deep galaxy surveys like the Keck Baryonic Structure
Survey (Steidel \etal\ 2004), zCOSMOS (e.g.\ Lilly \etal\ 2007) and
DEEP2 (Newman \etal\ 2012) have yielded a large sample of
spectroscopically identified galaxies $z=1-3$.  With a higher surface
density, galaxies identified in the background in these surveys in
principle provide an efficient means of densely mapping the CGM of
foreground galaxies identified in the same surveys (e.g.\ Adelberger
\etal\ 2005).  The caveat is, however, that galaxies are faint and
individual galaxy spectra do not provide a sufficient signal-to-noise
ratio (S/N) for identifying weak absorption features in the
foreground.  Therefore, it is necessary to average together a large
number of galaxy spectra in order to obtain sensitive constraints on
the CGM at these high redshifts (e.g.\ Steidel \etal\ 2010; Bordoloi
\etal\ 2011).  While using galaxy-galaxy pairs has substantially
improved the efficiency of probing the CGM using absorption
spectroscopy, it is worth noting that absorption constraints derived
from stacked spectra only give the mean properties of the gas averaged
over a sample of galaxies and do not constrain the incidence or
covering fraction of the gas.

With different empirical studies now available in the literature to
characterize the CGM in different epochs, it is interesting to examine
how the properties of the CGM have evolved with time.  Both galaxies
and the IGM exhibit very different properties at different redshifts.
For example, most galaxies at $z=2-3$ are seen actively forming stars
at higher rates than most galaxies at low redshift $z<0.5$ (e.g.\
Ouchi \etal\ 2009).  In addition, the ultraviolet background radiation
field is more than ten times higher at $z\approx 3$ than at $z\approx
0$ (e.g.\ Haardt \& Madau 2012).  Observations of the CGM as a
function of redshift may provide key insights into the physical
processes that drive the gas dynamics and chemical enrichment in
circumgalactic space.  This paper addresses the redshift evolution of
the CGM by comparing different absorption-line studies of halo gas
conducted at different redshifts.  Despite very different star
formation properties of the galaxies and different intergalactic
radiation fields at redshifts from $z\approx 2.2$ to $z\sim 0$, I show
that the extent and absorption strength of the CGM have not changed
over the past 11 billion years.  A $\Lambda$ cosmology, $\Omega_{\rm
  M}=0.3$ and $\Omega_\Lambda = 0.7$, with a Hubble constant $H_0 = 70
\ {\rm km} \ {\rm s}^{-1}\ {\rm Mpc}^{-1}$ is adopted throughout the
paper.

\begin{figure*}
\includegraphics[scale=0.5]{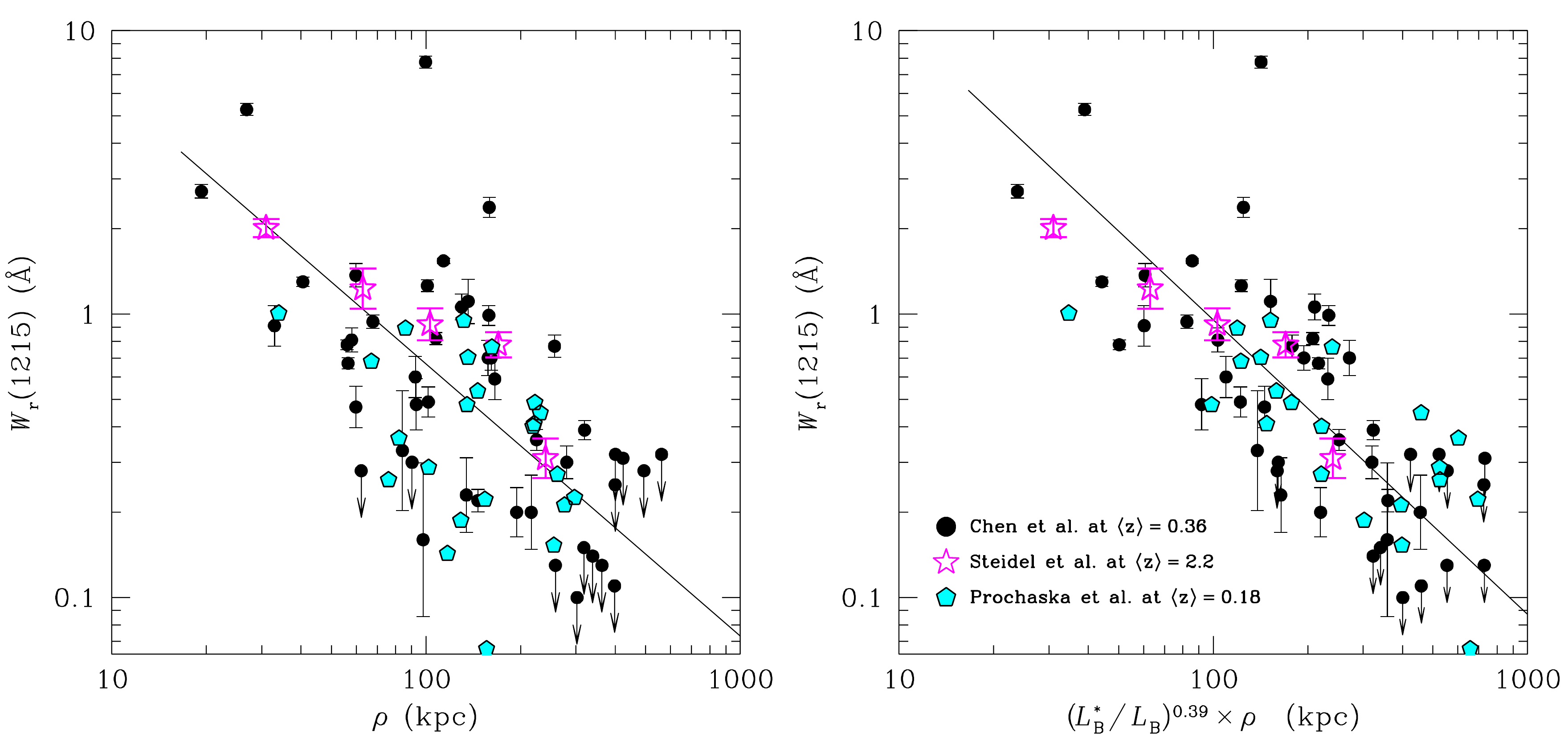}
\caption{Observed incidence and strength of \lya\ absorbers in the
  vicinities of galaxies over the redshift range from $z\sim 0$ to
  $z\sim 2.5$.  The {\it left} panel shows the rest-frame \lya\
  absorption equivalent width $W_r(1215)$ versus the projected
  distance of the absorbing galaxy $\rho$ in physical coordinates.
  Solid circles are galaxies from Chen \etal\ (1998, 2001a) with a
  median redshift of $\langle\,z\,\rangle=0.36$.  Pentagons are
  galaxies from Prochaska \etal\ (2011) with a median redshift of
  $\langle\,z\,\rangle=0.06$.  Star symbols are galaxies from Steidel
  \etal\ (2010) with a median redshift of $\langle\,z\,\rangle=2.2$.
  Errorbars indicate the 1-$\sigma$ measurement uncertainties of the
  absorber strength, while points with arrows indicate 3-$\sigma$
  upper limits for non-detections.  The solid line represents the
  power-law model of Chen \etal\ (2001a) that best characterizes the
  observed $W_r(1215)$ versus $\rho$ inverse-correlation for the
  $\langle\,z\,\rangle=0.36$ sample.  The {\it right} panel shows the
  rest-frame \lya\ absorption equivalent width $W_r(1215)$ versus
  projected distance normalized by $B$-band luminosity ($L_B$) for the
  low-redshift galaxy samples of Chen \etal\ and Prochaska \etal.  The
  luminosity scaling relation is adopted from Chen \etal\ for the
  $\langle\,z\,\rangle=0.36$ sample.  The best-fit power-law function
  (solid line) was determined based on a likelihood analysis that
  minimizes the scatter in the $W_r(1215)$ versus $\rho$ relation.
  The high-redshift measurements from Steidel \etal\ (star symbols)
  were made based on stacks of $\sim 100$ galaxies with a mean
  luminosity of roughly $L^*$ (D.\ Erb, private communication; see
  also Steidel \etal\ 2004).  Therefore, no scaling has been applied
  to these measurements in the right panel (star symbols).}
\end{figure*}

\section{The Galaxy Samples}

Five galaxy samples have been collected from the literature for
studying the CGM across cosmic time.  The galaxies are selected to
have available constraints on the properties of their halo gas from
absorption-line studies.  In addition, the discussion focuses on the
absorption species, H\,I, C\,IV, and Mg\,II for which measurements are
available from multiple epochs.  A brief description of the galaxy
samples and associated absorption-line measurements follows.

\subsection{The Chen \etal\ (1998, 2001ab)  sample at $\langle\,z\,\rangle=0.36$}

Chen \etal\ (1998, 2001a) presented a sample of 47 galaxies at
projected distances of $\rho \apl 300$ kpc\footnote{All numbers quoted
  in this paper have been corrected for the standard $\Lambda$
  cosmology.} (proper) from the lines of sight toward background QSOs
and obtained measurements (or upper limits) of the associated \lya\
absorption strength for individual galaxies in the spectra of the
background QSOs.  The galaxies in this sample span a broad range in
rest-frame $B$-band luminosities (from $\approx 0.03\,L_{B}^*$ to
$\approx 3\,L_{B}^*$ with a median of $\langle\,L_{B}\,\rangle =
0.6\,L_{B}^*$), a broad range in redshifts (from $z = 0.0752$ to
0.8920 with a median of $\langle\,z\,\rangle = 0.3567$), and a range
in impact parameter separations (from $\rho = 18$ to $250$ kpc with a
median of $\rho = 89$ kpc).  High spatial resolution optical images
obtained from the Hubble Space Telescope were available for
classifying the morphological types of these galaxies according to the
disk-to-bulge ratio ($D/B$) derived from a two-dimensional surface
brightness profile analysis.  The galaxies were found to exhibit a
broad range of optical morphologies, including bulge-dominated
elliptical/S0 ($D/B < 3$), earl-type spiral galaxies ($3 < D/B < 14$),
late-type disks ($D/B>14$), and galaxies with disturbed profiles that
cannot be characterized by a normal disk or bulge.

To constrain the incidence and strength of tenuous gas around the 47
galaxies, Chen \etal\ (1998, 2001a) searched for the associated \lya\
absorbers in the spectra of the background QSOs.  They detected
corresponding \lya\ absorbers at velocity separations $|\Delta\,v|\le
250$ \kms\ from 34 galaxies and placed 3-$\sigma$ upper limits to the
rest-frame \lya\ absorption equivalent width $W_r(1215)$ for the
remaining 13 galaxies.  The distribution of $W_r(1215)$ versus $\rho$
for this galaxy sample is shown in Figure 1 (solid circles).  These
authors report a strong $W_r(1215)$ versus $\rho$ anti-correlation
(left panel of Figure 1; see also Lanzetta \etal\ 1995) that depends
strongly on galaxy $B$-band luminosity, but not on galaxy morphology,
mean surface brightness, or redshift.

Using a likelihood analysis formulated to minimize the scatter in the
$W_r(1215)$ versus $\rho$ anti-correlation, Chen \etal\ (1998, 2001a)
concluded that the extent of tenuous gas around galaxies scales with
galaxy $B$-band luminosity as
$R_{\lya}/R_{\lya}^*=(L_B\,/\,L_{B}^*)^{0.39\pm 0.09}$ where
$R_{\lya}^*\approx 300$ kpc for $W_r(1215)=0.3$ \AA\ (right panel of
Figure 1) and that the mean gas covering fraction is $\apg 86$\%
within $\rho=R_{\lya}$.  The improved $W_r(1215)$ versus $\rho$
anti-correlation after taking into account the luminosity
normalization of the projected distances strongly supports that the
absorbers trace tenuous gas in individual halos surrounding the
galaxies rather than tenuous gas in galaxy groups or large-scale
filaments around the galaxy overdensities.

In a follow-up paper, Chen \etal\ (2001b) examined the incidence and
strength of C\,IV absorption features in the vicinities of 50
galaxies.  They detected C\,IV absorption features at velocity
separations $|\Delta\,v|\le 250$ \kms\ from 14 galaxies and placed
3-$\sigma$ upper limits to the rest-frame C\,IV\,$\lambda$\,1548
absorption equivalent width $W_r(1548)$ for the remaining 36 objects.
These authors found that while the $W_r(1548)$ versus $\rho$
distribution exhibits a large scatter, there appears to a distinct
boundary between C\,IV-absorbing and non-absorbing regions in galactic
halos.  Based on a likelihood analysis formulated to minimize the
scatter between detections and non-detections in the $W_r(1548)$
versus $\rho$ distribution, Chen \etal\ concluded that the extent of
C\,IV-enriched gas scales with galaxy $B$-band luminosity as $R_{\rm
  C\,IV}/R_{\rm C\,IV}^*=(L_B\,/\,L_{B}^*)^{0.5\pm 0.1}$ where $R_{\rm
  C\,IV}^*\approx 160$ kpc.  The luminosity-normalized $W_r(1548)$
versus $\rho$ distribution is presented in Figure 2 (solid circles).

Figure 2 shows that while the mean covering fraction of C\,IV-enriched
gas within $\rho=R_{\rm C\,IV}$ is high, $\apg 90$\%, around these
low-redshift galaxies, the scatter in the $W_r(1548)$ versus $\rho$
relation is also large.  To facilitate a comparison between different
studies, the galaxies within 160 kpc of the QSO sightlines in the
low-redshift study are divided into subsamples according to the impact
parameter bins adopted by Steidel et al. (2010).  The 1-$\sigma$
clipped mean and dispersion of $W_r(1548)$ in each subsample of Chen
\etal\ (2001b) are shown as the pentagon symbols and the associated
errorbars.  

Note that with a large scatter in $W_r(1548)$, the strongest absorbers
naturally dominate in calculating the mean value over a small sample
of $\sim 5$ galaxies per impact parameter bin (cf.\ $\sim 100$
galaxies per impact parameter bin in the stacks of Steidel \etal).
Applying a 1-$\sigma$ clipping criterion allows a more accurate
assessment of the mean intrinsic properties based on a rough estimate
of the underlying $W_r(1548)$ distribution.  As a result of the
clipping criterion, one galaxy at luminosity-normalized $\rho > 100$
kpc with an associated C\,IV absorber of $W_r(1548)\approx 1.7$ \AA\
is found to fall outside of the 1-$\sigma$ range and excluded from the
mean calculation.  This is the only galaxy excluded in the 1-$\sigma$
clipped mean calculation of the Chen \etal\ sample.


\begin{figure}
\includegraphics[scale=0.45]{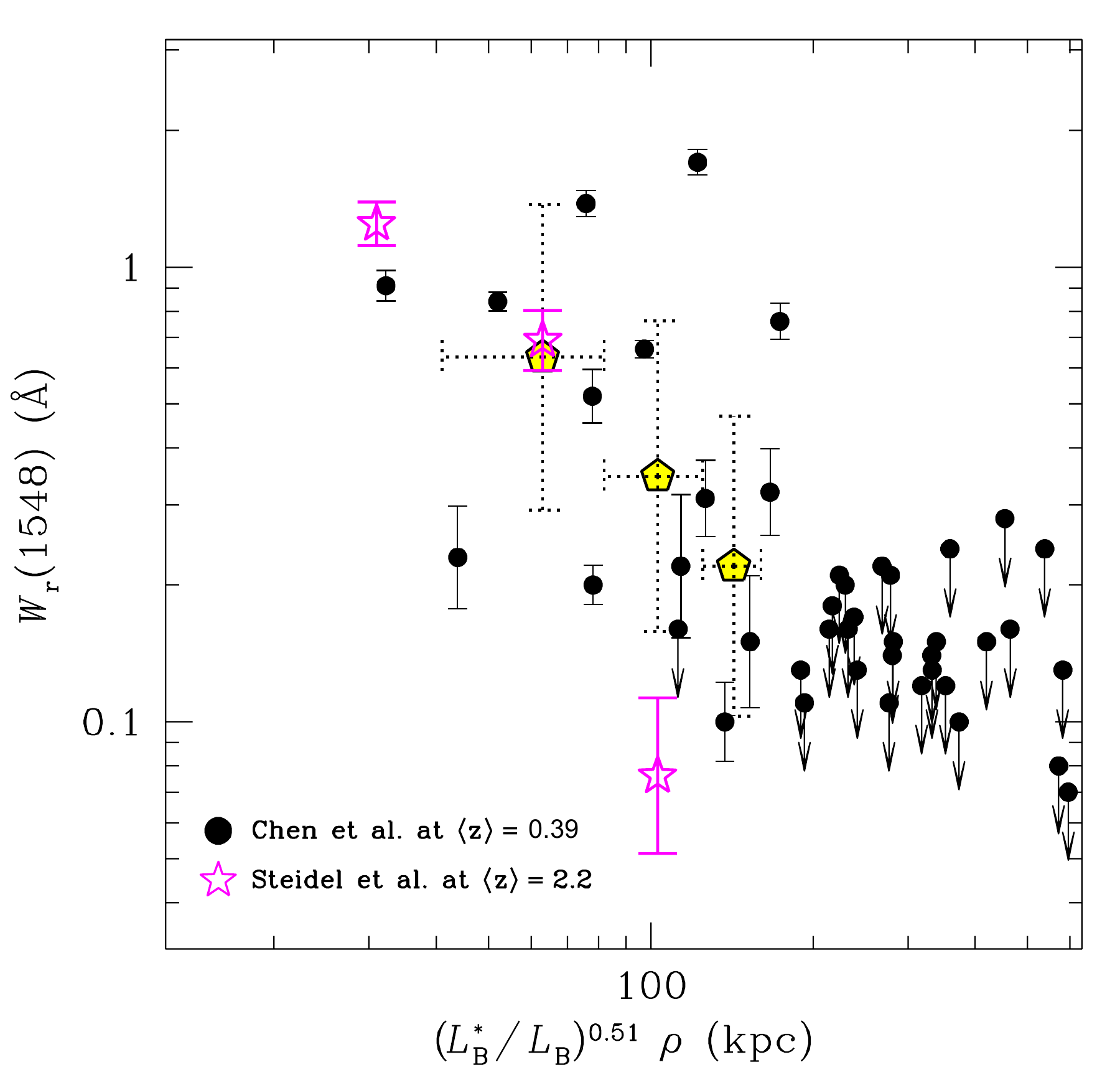}
\caption{Observed incidence and strength of C\,IV absorbers in the
  vicinities of galaxies from two different redshift ranges, $z\apl
  0.5$ and $z\sim 2.5$.  Solid circles are galaxies from Chen \etal\
  (2001b) with a median redshift of $\langle\,z\,\rangle=0.39$.
  Errorbars indicate the 1-$\sigma$ measurement uncertainties of the
  absorber strength, while points with arrows indicate 3-$\sigma$
  upper limits for non-detections.  Impact parameters of these
  galaxies have been normalized by their $B$-band luminosity according
  to the best-fit scaling relation of Chen \etal\ (2001b) that
  minimizes the scatter between detections and non-detections.  Star
  symbols are galaxies from Steidel \etal\ (2010) with a median
  redshift of $\langle\,z\,\rangle=2.2$ and a mean luminosity of
  roughly $L^*$.  Because the absorption spectra of Steidel \etal\ did
  not have sufficient resolution to resolve the C\,IV doublet, these
  authors reported the combined absorption equivalent width of the
  doublets.  The star symbols and associated errorbars in this panel
  indicate the estimated absorption strength of the
  C\,IV\,$\lambda$\,1548 component and its associated uncertainties
  due to measurement errors and uncertain doublet ratios (in cases of
  saturated lines).  To facilitate the comparison between low- and
  high-redshift samples, the galaxies within 160 kpc of the QSO
  sightlines in the low-redshift study are divided into subsamples
  according to the impact parameter bins adopted by Steidel et
  al. (2010).  The 1-$\sigma$ clipped mean and dispersion of
  $W_r(1548)$ in each subsample of Chen \etal\ (2001b) are shown as
  the pentagon symbols and the associated errorbars.}
\end{figure}

\subsection{The Chen \etal\ (2010a)  sample at $\langle\,z\,\rangle=0.25$}

Using a sample of 94 galaxies spectroscopically identified at
redshifts between $z=0.11$ and $z=0.49$ with a median redshift of
$\langle z\rangle = 0.24$ and located at $\rho\apl 170$ kpc from the
line of sight of a background QSO, Chen \etal\ (2010a) examined the
incidence and strength of Mg\,II absorption features around galaxies.
The galaxies in this sample span a broad range in rest-frame $B$-band
luminosities, from $\approx 0.04\,L_{B}^*$ to $\approx 4\,L_{B}^*$
with a median of $\langle\,L_{B}\,\rangle = 0.8\,L_{B}^*$, and a broad
range in rest-frame $B_{AB}-R_{AB}$ colors from $B_{AB}-R_{AB}\approx
0$ to $B_{AB}-R_{AB}\approx 1.5$.  In a follow-up study, Chen \etal\
(2010b) estimated the total stellar mass and mean star formation rate
average over the last 10 Myr by analyzing the broad-band spectral
energy distribution of these galaxies.  The galaxies are found to span
a stellar mass range from $\log\,M_{\rm star}=8.5$ to $\log\,M_{\rm
  star}=11.6$ with a median of $\langle\,\log\,M_{\rm
  star}\rangle=10.3$ and a range in specific star formation rate from
$\log\,{\rm sSFR}=-11.7$ to $\log\,{\rm sSFR}=-8.7$ with a median of
$\langle\,\log\,{\rm sSFR}\rangle=-10.1$.

Excluding galaxies with known neighbors at $\rho< 200$ kpc, Chen
\etal\ (2010a) detected corresponding Mg\,II absorbers at velocity
separations $|\Delta\,v|\apl 300$ \kms\ 47 galaxies and placed
2-$\sigma$ upper limits to the rest-frame Mg\,II\,$\lambda$\,2796
absorption equivalent width $W_r(2796)$ for the remaining 24 galaxies.
Similar to the study of extended \lya\ absorbing gas, these authors
report a strong $W_r(2796)$ versus $\rho$ anti-correlation that
depends strongly on galaxy $B$-band luminosity or stellar mass, but
not on galaxy colors or redshift (Chen \etal\ 2010a,b).  Based on a
likelihood analysis formulated to minimize the scatter in the
$W_r(2796)$ versus $\rho$ anti-correlation, the authors concluded that
the extent of Mg\,II-enriched gas around galaxies scales with galaxy
$B$-band luminosity as $R_{\rm Mg\,II}/R_{\rm
  Mg\,II}^*=(L_B\,/\,L_{B}^*)^{0.35\pm 0.03}$ where $R_{\rm
  Mg\,II}^*\approx 107$ kpc and that the mean gas covering fraction is
$\approx 80$\% within $\rho=R_{\rm Mg\,II}$ for absorbers of
$W_r(2796) \ge 0.1$ \AA.

The luminosity-normalized $W_r(2796)$ versus $\rho$ distribution is
presented in Figure 3 (solid circles).  Outliers identified based on a
3-$\sigma$ clipping criterion are highlight in dotted circles.
Galaxies with close neighbors are indicated by open symbols.

\subsection{The Steidel \etal\ (2010) sample at $\langle\,z\,\rangle=2.2$}

Steidel \etal\ (2010) examined the CGM at $z>2$ using a sample of 512
close galaxy pairs separated by $\apl 15''$.  At the median redshift
of the galaxy pair sample $\langle\,z\rangle=2.2$, an angular
separation of $15''$ corresponds to projected separations of $\apl
124$ kpc.  Galaxies in each pair are found at cosmologically distinct
redshifts with line-of-sight velocity separation $>3000$ \kms, and
therefore the background galaxy serves as a random probe of unseen gas
associated with the foreground galaxy (see also Adelberger \etal\
2005). ${\cal R}$-band magnitudes of these galaxies range from ${\cal
  R}=22$ to ${\cal R}=25.5$ with a median $\langle\,{\cal
  R}\,\rangle\approx 24.3$ (e.g.\ Steidel \etal\ 2004), corresponding
to roughly $L^*$ at $z\sim 2.3$ (e.g.\ Reddy \& Steidel 2009).  The
mean stellar mass of these galaxies is found to be around
$\langle\,\log\,M_{\rm star}\rangle\approx 10.3$ (Shapley \etal\
2005).  Many of these galaxies are actively forming stars, showing
relatively luminous UV fluxes and strong blueshifted self-absorption
in the UV ISM lines (e.g.\ Shapley \etal\ 2003; Steidel \etal\ 2004;
Adelberger \etal\ 2005).

To map the CGM of these star-forming galaxies, Steidel \etal\ first
divided the galaxy sample in three impact parameter bins, $\rho\le 41$
kpc, $\rho=41-83$ kpc, and $\rho=83-124$ kpc, and in each bin they
formed an average spectrum of the background galaxies after shifting
the individual spectra to the rest frame of their foreground members.
The mean spectra were of high $S/N$ and revealed several absorption
features imprinted by the CGM of the foreground galaxies.  These
include hydrogen $\lya\,\lambda\,1215$,
C\,IV\,$\lambda\lambda\,1548,1550$, and Si\,II\,$\lambda\,1260$,
although the relatively low resolution of the spectra did not allow
the authors to resolve the C\,IV doublet.

Measurements of the absorber strength versus impact separation from
Steidel \etal\ (2010) are included in Figures 1 through 3 (star
symbols) for direct comparisons with observations of the CGM at lower
redshifts.  Note that the $W_r(1215)$ measurements at $\rho > 125$ kpc
were made using stacks of high $S/N$ spectra of background QSOs in the
rest frame of 21 foreground galaxies.  In addition, the original C\,IV
absorption strength reported by Steidel \etal\ (2010) include both
components of the doublet.  The values and associated uncertainties of
$W_r(1549)$ for the $z\approx 2.2$ galaxies in Figure 2 are estimated
based on a range of possible C\,IV doublet ratio from 2:1 to 1:1 due
to line saturation.  Finally, Mg\,II absorption doublets are
redshifted into the near-infrared spectral window at $z\apg 2.2$, and
are therefore missed in the available optical spectra of Steidel
\etal\ (2010).  However, the Si$^+$ and $Mg^+$ ions have comparable
ionization potentials and are observed to coexist in a broad range of
astrophysical environment (e.g.\ Sembach \& Savage 1996;; Ganguly
\etal\ 2005; Churchill \etal\ 2000).  It is therefore justified to
infer the associated Mg\,II absorption strength $W_r(2796)$ of the
$z\approx 2.2$ galaxies based on known absorption strength of
Si\,II\,$\lambda\,1260$, $W_r(1260)$ from Steidel et al.  Based on the
observations of Churchill \etal\ (2000), I estimate that
$W_r(2796)\approx 1.5\,W_r(1260)$.  The inferred $W_r(2796)$ of the
$z\approx 2.2$ galaxies are shown in Figure 3 as star symbols.

\begin{figure}
\includegraphics[scale=0.47]{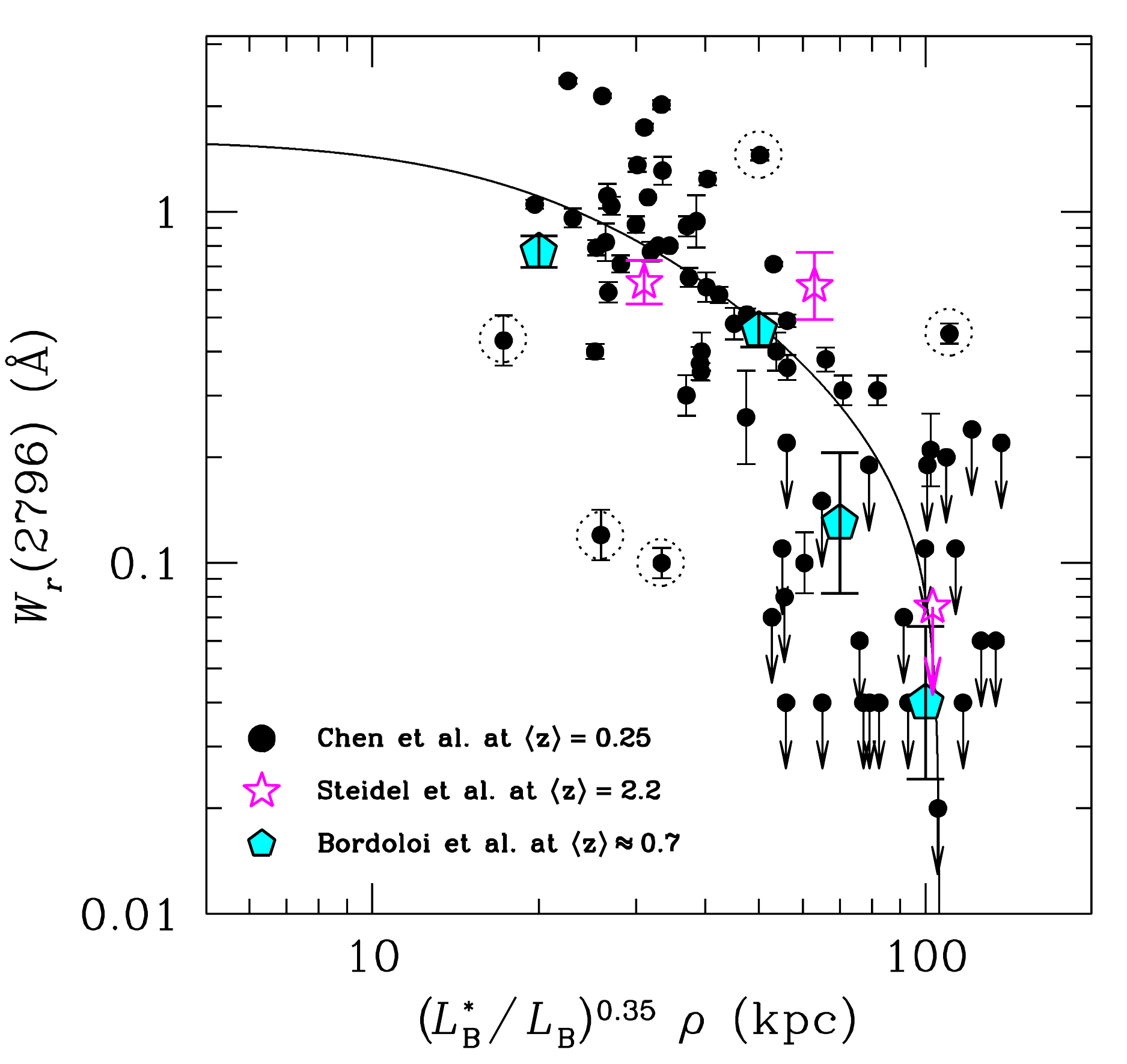}
\caption{Observed incidence and strength of Mg\,II absorbers in the
  vicinities of galaxies from three different epochs, $z\apl 0.5$,
  $z=0.5-0.9$, and $z\sim 2.5$.  Solid circles are galaxies from Chen
  \etal\ (2010a) with a median redshift of $\langle\,z\,\rangle=0.24$.
  Errorbars indicate the 1-$\sigma$ measurement uncertainties of the
  absorber strength, while points with arrows indicate 2-$\sigma$
  upper limits for non-detections.  Impact parameters of these
  galaxies have been normalized by their $B$-band luminosity according
  to the best-fit scaling relation of Chen \etal\ (2010a) from a
  likelihood analysis that strengthens the $W_r(2796)$ versus $\rho$
  inverse correlation.  The likelihood analysis excluded outliers
  (points with dotted circles) identified using a 3-$\sigma$ clipping
  criterion.  The solid curve represents the best-fit model that
  follows a $r^{-2}$ density profile of maximum extent $R_{\rm
    Mg\,II}$ (see Chen \etal\ 2010a for details).  Pentagons are for
  blue galaxies at $z=0.5-0.9$ from Bordoloi \etal\ (2011) with
  $I_{\rm AB}\le 22.5$ and a median redshift of
  $\langle\,z\,\rangle\approx 0.7$.  At $z=0.7$, $I_{\rm AB}=22.5$
  corresponds to $L^*$.  Therefore the measurements of Bordoloi \etal\
  have not been corrected for luminosity differences.  Star symbols
  are the expected $W_r(2796)$ for $\sim\,L_*$ galaxies at
  $\langle\,z\,\rangle=2.2$, based on the rest-frame absorption
  equivalent width of the Si\,II\,$\lambda$\,1260 transition
  $W_r(1260)$ from Steidel \etal\ (2010).  At $z\apg 2.2$, Mg\,II
  absorption doublets are redshifted into the near-infrared spectral
  window and not observed by optical spectra.  The expected
  $W_r(2796)$ is calculated based on the empirical correlation,
  $W_r(2796)\approx 1.5\,W_r(1260)$, extracted from observations of
  Churchill \etal\ (2000).}
\end{figure}

\subsection{The Prochaska \etal\ (2011)  sample at $\langle\,z\,\rangle=0.18$}

Prochaska \etal\ (2011) conducted a galaxy survey in the fields of 14
QSOs at $z_{\rm QSO}=0.06-0.57$ for which UV absorption spectra were
available for searching \lya\ and O\,VI absorption features associated
with the low-redshift CGM.  The survey reached $>60$\% completeness
for galaxies brighter than $R=19.5$ mag within $5'$ radius of the QSO
lines of sight.  The galaxies in this sample span a broad range in
rest-frame $B$-band luminosities, from $\approx 0.01\,L_{B}^*$ to
$\approx 5\,L_{B}^*$ with a median of $\langle\,L_{B}\,\rangle =
0.3\,L_{B}^*$, and a broad range in redshifts, from $z = 0.005$ to 0.5
with a median of $\langle\,z\,\rangle = 0.18$.  

Considering only galaxies at $\rho \le 300$ kpc from the lines of
sight toward background QSOs, the authors examined the incidence and
strength of \lya\ in the vicinities of these galaxies.  Crosses in
Figure 1 represent the results from Prochaska \etal\ (2011) with
(right panel) and without (left panel) the luminosity scaling.

\subsection{The Bordoloi \etal\ (2011) sample at $\langle\,z\,\rangle\approx 0.7$}

The zCOSMOS team (Lilly \etal\ 2007) has produced a large sample of
spectroscopically identified galaxies with $I_{\rm AB}\le 22.5$ mag at
$z=0-1.4$ over the 2 deg$^2$ sky area covered by the COSMOS project
(Scoville \etal\ 2007).  Additional spectroscopic data have also been
obtained in the follow-up zCOSMOS-deep effort that targeted fainter
galaxies ($B<25.5$ mag) in the inner 1 deg$^2$ region of the COSMOS
field with photometric redshifts $z_{\rm phot}>1$.  Taking advantage
of the existing large galaxy survey data, Bordoloi \etal\ (2011)
adopted the approach of Steidel \etal\ (2010) and examined the
properties of the metal-enriched CGM at intermediate redshifts based
on the incidence and strength of Mg\,II absorbers in the vicinities of
zCOSMOS galaxies.

Specifically, Bordoloi \etal\ first identified foreground galaxies
with redshift accurately known at $z=0.5-0.9$ from the zCOSMOS-bright
sample.  These authors then identified background galaxies with
spectra available from zCOSMOS-deep.  Although the zCOSMOS-deep
spectra did not have sufficient quality for measuring the redshifts
for some of these objects, the $z_{\rm phot}>1$ selection criterion
ensures that the majority of these objects are at $z>1$ and therefore
their optical spectra are still useful for searching for intervening
absorption features at $z<1$.  Cross-correlating the two galaxy
samples, Bordoloi \etal\ identified a sample of $\approx 3900$
galaxies at $z=0.5-0.9$ and located within $\rho\le 200$ kpc of a
background galaxy.  The spectra of the background galaxies were
co-added in the rest frame of the foreground objects.  The spectral
coverage allowed these authors to search for Mg\,II absorption
features associated with the foreground galaxies and to examine the
properties of the metal-enriched CGM based on the observed spatial
distribution of $W_r(2796)$.  The observed $W_r(2796)$ versus $\rho$
distribution for blue galaxies from Bordoloi \etal\ (2011) is shown as
cyan crosses in Figure 3.

Note that the consideration of only blue galaxies from Bordoloi \etal\
is justified here by their mean stellar mass range
$\langle\,\log\,M_{\rm star}\rangle=10-10.3$, which is comparable to
galaxies in the Chen \etal\ sample (\S\ 2.2) and Steidel \etal\ (\S\
2.3) samples shown in the same figure.  The red galaxies in the
Bordoloi \etal\ sample are more massive with $\langle\,\log\,M_{\rm
  star}\rangle\approx 10.7$.  In addition, Chen \etal\ (2010b) showed
that the majority of galaxies in the Chen \etal\ (2010a) sample fall
on the blue sequence.  Few galaxies exhibit colors that are consistent
with red galaxies in the field.  Most of their red galaxies also
turned out to be in a group environment and were excluded from their
analysis.  The incidence and strength of Mg\,II absorbers around
massive red galaxies have been found suppressed relative to younger
and lower mass objects (e.g.\ Gauthier \etal\ 2010, Gauthier \& Chen
2011; Bordoloi \etal\ 2011).  The red sample of Bordoloi et al.\ is
therefore not considered in this study.

Finally, the exact distribution of $B$-band luminosity in Bordoloi
\etal\ sample is not known.  However, at $z=0.7$, the limiting
magnitude $I_{\rm AB}=22.5$ of the zCOSMOS sample corresponds to $L^*$
(see e.g.\ Faber \etal\ 2007).  Therefore the measurements of Bordoloi
\etal\ have not been corrected for luminosity differences in Figure 3.

\section{Discussion}

The collection of five absorption-line surveys in the vicinities of
distant galaxies offers a unique opportunity to examine for the first
time how the properties of the CGM have evolved with redshift.  It is
immediately clear from Figures 1, 2, and 3 that the observed absorber
strength versus luminosity-normalized projected distance does not vary
from $z\approx 2.2$ to $z\sim 0$ for ionized gas probed by either
hydrogen, C\,IV, or Mg\,II transitions.  To assess the statistical
significance of the observed lack of redshift evolution, I perform a
$\chi^2$ test to compare the absorption properties of the CGM at
different redshifts.  The results are summarized in Table 1.

Specifically, I first adopt the best-fit power-law function of Chen
\etal\ (2001a) at $\langle z\rangle=0.36$ as the fiducial model.  Then
I estimate the probability of the observed radial distribution of
hydrogen absorption around $\langle z\rangle=2.2$ galaxies being drawn
from the fiducial model.  I find that the probability of the two
samples sharing the same parent distribution is $>99$\%.  Next, I
repeat the same approach for Mg\,II absorbing gas, adopting the
best-fit model profile of Chen \etal\ (2010a) at $\langle z\rangle =
0.25$ as the fiducial model and estimating the probability of each of
the Bordoloi \etal\ (2011) and Steidel \etal\ (2010) samples being
drawn from the fiducial model.  I find that the probability of
galaxies possessing the same radial absorption profile of
low-ionization species since $z=2.2$ is better than 97\%.  The result
for C\,IV absorbing gas is more uncertain, because fewer galaxies were
found at luminosity-normalized $\rho<160$ kpc in the low-redshift
sample.  Owing to the observed large scatter in Figure 2, no analytic
model presents a good fit to the data.  The $\chi^2$ test is therefore
performed based on comparisons of the binned data (pentagons versus
star symbols in Figure 2).  This exercise also yields a high
probability of 80\% that the low- and high-redshift samples share
similar C\,IV absorption properties in the CGM.  In summary, the
$\chi^2$ test therefore supports the conclusion that the CGM exhibits
little change in the absorption properties over the redshift range
from $z=0.2$ to $z=2.2$.

\begin{footnotesize}
\begin{table}
  \footnotesize
  \centering
  \begin{minipage}{160mm}
    \caption{Summary of the Galaxy Samples for CGM Studies}
    \begin{tabular}{@{}lccc@{}}
      \hline
      \multicolumn{1}{c}{Sample}  & Probe  &  \multicolumn{1}{c}{$\langle z\rangle$}  & \multicolumn{1}{c}{$\chi^2$ test} \\
      \hline
      \hline
      \multicolumn{4}{c}{hydrogen absorption} \\
      \hline
      Chen \etal\ (1998, 2001a)  & QSO-galaxy    & 0.36 &  \\
      Steidel \etal\ (2010)      & Galaxy-galaxy & 2.20 & 99\% \\
      Prochaska \etal\ (2011)    & QSO-galaxy    & 0.18 & 99\% \\
      \hline
      \multicolumn{4}{c}{C\,IV absorption} \\
      \hline
      Chen \etal\ (2001b)        & QSO-galaxy    & 0.39 &  \\
      Steidel \etal\ (2010)      & Galaxy-galaxy & 2.20 & 80\% \\
      \hline
      \multicolumn{4}{c}{Mg\,II absorption} \\
      \hline
      Chen \etal\ (2010a)        & QSO-galaxy    & 0.25 &  \\
      Steidel \etal\ (2010)$^a$      & Galaxy-galaxy & 2.20 & 97\% \\
      Bordoloi \etal\ (2011)     & Galaxy-galaxy & $\approx 0.7$ & 99\%  \\
      \hline
      \multicolumn{4}{l}{$^a$As described in the text, the Mg\,II absorption strength here}\\ 
      \multicolumn{4}{l}{\ is inferred from the reported Si\,II 1260 transition.}
    \label{pair_table}
  \end{tabular}
\end{minipage}
\end{table}
\end{footnotesize}

Note that {\it the lack of variation applies to both mean absorption
  equivalent width and the spatial extent of the CGM traced by either
  hydrogen or heavy ions such as C$^{3+}$ and Mg$^+$.}  The absorption
features adopted for the comparison are all strong transitions and are
likely saturated.  Although measurements of the rest-frame absorption
equivalent width for saturated lines do not constrain the underlying
total gas column densities, they constrain the mean number of
absorbing clumps and gas kinematics along the lines of sight (e.g.\
Petitjean \& Bergeron 1990; Prochter \etal\ 2006).  It is therefore
important to keep in mind that the agreement in the mean absorption
equivalent width between the CGM at low and high redshifts reflects
primarily similar kinematics and volume filling factor of gaseous
clumps, not total gas density, in galactic halos.  But whether or not
the absorption lines are saturated does not change the conclusion that
the spatial extent ($\approx 100$ kpc) of the chemically CGM has
varied little with time.  Here I consider possible systematics and
implications.

\subsection{Point sources versus extended background light}

A fundamental difference between CGM studies at low and high redshifts
is the use of point-like QSOs versus spatially extended galaxies as
the background light source.  The UV continuum emission regions of
QSOs are $<$ 1 pc (e.g.\ Blackburne \etal\ 2012), while galaxies at
$z>1$ have a typical half-light radius of $\sim 2-3$ kpc (e.g.\ Law
\etal\ 2007).  If the foreground gaseous clouds do not fully cover the
background galaxies, then the absorption equivalent width would appear
to be weaker than what is seen in the spectrum of a background QSO
when the QSO light is fully covered by the clouds.  Comparisons
between measurements made along QSO and galaxy sightlines therefore
need to take into account the effect of possible partial covering
fractions.

A key result in the low-redshift surveys summarized in \S\ 2 is that
the covering fraction of extended gas around galaxies is nearly unity.
Specifically, the mean covering fraction of tenuous gas probed by
\lya\ is $\apg 86$\% within $\rho=R_{\lya}$; the mean covering
fraction of C\,IV-enriched gas within $\rho=R_{\rm C\,IV}$ is $\apg
90$\%; and the mean covering fraction of Mg\,II-enriched gas within
$\rho=R_{\rm Mg\,II}$ is $\apg 80$\%.  Accounting for possible
partial covering fractions is therefore unlikely to change the
measurements of Bordoloi \etal\ (2011) and Steidel \etal\ (2010) by
more than 20\%.  A high gas covering fraction is confirmed in CGM
surveys at $z\apg 1$ using QSO probes by Lovegrove \& Simcoe (2011)
and by Adelberger \etal\ (2005), albeit with large uncertainties due
to a small sample size.

\subsection{Random velocity offsets between galaxies and absorbers}

Random velocity offsets between absorbing gas and the systemic motion
of the absorbing galaxy can in principle result in a suppressed
absorption signal in stacked spectra (e.g.\ Ellison \etal\ 2000).  If
this is a dominant effect, then the measurements of Bordoloi \etal\
(2011) and Steidel \etal\ (2010) should be treated as lower limits.

The velocity separation between galaxies and absorbers identified
along QSO lines of sight is well characterized by a Gaussian
distribution function (e.g.\ Lanzetta \etal\ 1997; Chen \etal\ 2010a).
Specifically, the velocity dispersion between Mg\,II absorbers and
their absorbing galaxies is found to be $\sigma_v\approx 137$ \kms\
(Chen \etal\ 2010a).  To assess the magnitude of such systematic bias,
I perform a Monte Carlo simulation that creates stacks of 50 to 100
individual mocked absorption spectra with a spectral resolution of
${\rm FWHM}\approx 350$ \kms\ (comparable to the spectral quality for
the high-redshift CGM studies) and varying $S/N$ from $S/N=5$ to
$S/N=10$.  A synthetic absorption feature is randomly placed in each
mocked spectrum following a Gaussian distribution function around a
fiducial position.  I experiment with different widths of the Gaussian
distribution function, ranging from $\hat\sigma_v=50$ to
$\hat\sigma_v=600$ \kms\ and measure the absorption equivalent widths
in the stacked spectra.  The results of the Monte Carlo simulation
study shows that for 0.1-\AA\ absorbers one can recover the full
absorption equivalent width at $\hat\sigma_v\apl 200$ \kms.  At
$\hat\sigma_v\approx 300$ \kms, the absorption equivalent width may be
underestimated by $\approx 20$\%.  Given that the velocity dispersion
between galaxies and their absorbers is less than 200 \kms, I find it
unlikely that the absorption equivalent width measurements in stacked
spectra are significantly underestimated due to random velocity
offsets between gas and galaxies.

\subsection{The nature of $L^*$ galaxies at different redshifts}

The spatial profiles of the CGM absorption shown in Figures 1 through
3 are based on luminosity-normalized projected distances.  This is
motivated by the expectation that more massive galaxies possess more
extended gas (e.g.\ Mo \& Miralda-Escude 1996).  Because the
observations are calibrated with respect to $L^*$-type galaxies in
different epochs, it is also necessary to examine the intrinsic
properties of $L^*$ galaxies at different redshifts before attempting
to understand the lack of variation in the observed spatial
distributions of absorbers around distant galaxies.

Using an abundance matching approach, several authors have estimated
the mean halo mass for galaxies of different luminosity/mass at
different epochs (e.g.\ Zheng \etal\ 2007; Conroy \etal\ 2008).  At
low redshifts, $L^*$ galaxies are found to reside on average in
$M_h\sim 10^{12}\,{\rm M}_\odot$ halos based on a luminosity-dependent
mapping between galaxies and dark matter halos (Zheng \etal\ 2007).
Likewise, Conroy \etal\ (2008) found that the majority of
spectroscopically identified star-forming galaxies at $z\sim 2$ reside
in halos of $M_h\sim 10^{12}\,{\rm M}_\odot$.  Therefore, $L^*$
galaxies appear to reside in halos of comparable mass at these
different redshifts.  

This conclusion remains if one infers the mean halo mass of these
galaxies based on the mean observed stellar masses.  At both low and
high redshifts, these galaxies are typically found to contain $M_{\rm
  star} \sim 2\times 10^{10}\,{\rm M}_\odot$, which leads to a typical
dark matter halo mass of $M_h\sim 10^{12}\,{\rm M}_\odot$ (e.g.\
Conroy \& Wechsler 2009; Moster \etal\ 2010; Behroozi \etal\ 2010).
Together the results of this exercise confirm that the comparisons
shown in Figures 1 through 3, after the normalization to $L^*$
galaxies, are for halos of comparable mass and therefore support the
conclusion that the spatial extent and mean absorption strength of the
CGM in halos of comparable mass have changed little since $z\sim 2$.

\subsection{Implications}

The consistent spatial profiles of different species in the CGM from
$z\approx 2.2$ to $z\sim 0$ are difficult to interpret, whether the
observed absorbers originate in starburst driven outflows (e.g.\
Martin \& Bouch\'e 2009; Weiner \etal\ 2009; Rubin \etal\ 2010) or in
infalling clouds (e.g.\ Heitsch \& Putman 2009).  In particular,
galaxies at $z = 2 - 3$ are seen actively forming stars and super
galactic winds may be effective in replenishing the CGM with
metal-enriched gas (e.g.\ Murray \etal\ 2011).  In contrast, galaxies
at $z< 0.5$ are more quiescent with a diverse star formation history
and super galactic winds are not expected to be the primary
explanation for the high incidence and covering fraction of
metal-enriched, cool halo gas out to 100 kpc.

At the same time, the intergalactic ultraviolet radiation field has
declined by more than a factor of ten from $z=2-3$ to $z\sim 0$.  Cool
clouds are expected to form at larger galactic distances at lower
redshifts (e.g.\ Mo \& Miralda-Escude 1996).  If the observed
absorbers originate in infalling cool clouds, then the spatial profile
is expected to be more extended at lower redshift, which also
disagrees with the trend shown in \S\ 2.

In summary, comparisons between different absorption-line surveys
conducted in the vicinities of galaxies at different redshifts have
revealed a puzzling finding.  Despite very different star formation
properties of the galaxies considered in these separate studies and
different intergalactic radiation fields at redshifts between
$z\approx 2.2$ and $z\sim 0$, the spatial extent and mean absorption
strength of the CGM around galaxies of comparable mass appear to be
unchanged over this cosmic time interval.  The observed lack of
variation in the CGM properties at different redshifts provides a
critical test of theoretical models of gas flows around galaxies.

\section*{Acknowledgments}

It is a pleasure to thank Nick Gnedin, Sean Johnson, Andrey Kravtsov,
Lynn Matthews, and Michael Rauch for important discussions.  I thank
Sean Johnson, Lynn Matthews and Michael Rauch for helpful comments on
an earlier version of this paper.


\label{lastpage}

\end{document}